\documentclass{article}[12 pt]
\usepackage{amsmath,amssymb,cite,epsfig,mathrsfs}
\begin{document}

\title{\bf{Finding Schwarzschild metric component $g_{rr}$ and FLRW's k without solving the Einstein's equation, rather by a synergistic matching between geometric results enfranchised by Newtonian gravity} }

\author{Eduardo I. Guendelman$^{1, 2, 3}$\thanks{e-mail: guendel@bgu.ac.il},
Avi Rabinowitz$^{1}$\thanks{e-mail: air1@nyu.edu }, Arka P. Banik$^{4}$\thanks{e-mail: apb3hf@mail.missouri.edu}\\
{\small $^1$ Ben Gurion University of the Negev, Department of Physics, Beer-Sheva, Israel} \\
{\small $^2$ Bahamas Advanced Study Institute and Conferences, 4A Ocean Heights, Hill View Circle,
 Stella Maris, Long Island, The Bahamas}\\ 
{\small $^3$ Frankfurt Institute for Advanced Studies, Giersch Science Center, Campus Riedberg,
Frankfurt am Main, Germany}\\
{\small $^4$ Department of Physics and Astronomy, University of Missouri - Columbia , Columbia, MO, USA}}

\maketitle

\begin{abstract}

As  is well known, some aspects of General Relativity and Cosmology can be reproduced without even using Einstein's equation. As an illustration, the $0-0$ component of the Schwarzschild space can be obtained by the requirement that the geodesic of slowly moving particles match the Newtonian equation. Given this result, we shall show here that the remaining component ($g_{rr}$) can be obtained by requiring that the inside of a Newtonian ball of dust matched at a free falling radius with the external space of unspecified type. This matching determines the external space to be of Schwarzschild type. By this, it is also possible to determine that the constant of integration that appears in the Newtonian Cosmology, coincides with the spatial curvature of the FLRW metric. All we assumed was some classical boundary conditions and basic assumptions.
\end{abstract}

\section{Introduction to our approach}

Newtonian gravity, as we know, can be seen as the effect of the $g_{00}$ metric component arising from the Schwarzschild metric. In the Newtonian regime where it can be seen as solely dependent on $g_{00}$. Newtonian theory on its own can therefore not make statements about spatial curvature, for example it cannot specify $g_{rr}$. However, there is a way partly around this restriction. Using Newtonian theory one can create a Newtonian cosmology, specifically a field equation governing the expansion or collapse of a cloud of dust (the universe). Of course, there is no theoretical justification a priori for assuming this has any validity within the General Relativity (GR) context. However as it turns out, using general arguments in General Relativity (Equivalence Principle) one arrive as at the conclusion that this aspect of Newtonian theory can indeed be legitimately utilized, after of course re-interpreting appropriately the dynamics as spatial expansion (rather than collapse or explosion) and suitably reinterpreting the coordinates and constants in the equation.
 
\medskip
Since, Newtonian theory does not contain spatial curvature however, there is no possibility of it emerging legitimately in a cosmological model based solely on Newtonian gravity. However, once we reinterpret the Newtonian cosmology field equation as a GR field equation, i.e. as one of the Einstein equations, this is not any more theoretically impossible and we will indeed see how we can obtain spatial curvature information.

\medskip
For the case of a particle moving around a central object of mass M, using
the Newtonian second law, one can replace $F =\frac{GMm}{r^2}$ by $a = \frac{GM}{r}$. This embodies the Equivalence Principle,  since the value of a clearly
does not depend on the object affected. As such, gravity can be considered
as " an acceleration field ". Considering for simplicity just radial motion,
and then calling $r(t)$ of a free fall particle its "worldline",
According to the EP a free fall particle is locally-inertial and taking
this fully seriously as Einstein did, leads us to conclude that a free
fall particle has a geodesic worldline, which means that that the
spacetime must be curved. Also, taking seriously the fact that inertial forces are indistinguishable
from gravitational forces, one concludes that the Earth-stationary
observer is accelerated, and so its worldline is necessarily curved.
Consequently, the free fall particle has a geodesic worldline and the
stationary observer watching it 'fall' has a curved worldine, the opposite
of what is reported by Newtonian gravity theory. This "switched perspective" is based on Einstein's understanding that if
phenomena are indistinguishable, they are identical, thus leading him to
say, based on the EP, that free fall is not just locally
indistinguishable from inertial, it is inertial. The connection is the
geometric property which grants a manifold the ability to define
'straightness', i.e. geodesity. Geodesics are by definition straight, one
can approach then this in two ways,
Newton-Cartan theory is a formal framework presenting this idea of gravity
being a (non-trivial) connection on spacetime, giving rise to non-zero
Riemann curvature. That one must only 'switch perspectives' to get from
EP-based Newtonian gravity to Newton-Cartan theory indicates how similar
they are, and perhaps explains why some results of Newtonian theory can
easily be adapted to become results of a curved spacetime theory of
gravity, such as Einstein's general relativity \cite{16}.

\medskip
Einstein's theory differs from Newton-Cartan theory mostly in that an
additional geometric structure is imposed on spacetime - a metric. In
ordinary life we can see that there is obviously a metric on space, and a
metric on time, but we certainly have no direct intuition of 'spacetime',
let alone a metric on it. Newton-Cartan theory adds the notion of
spacetime, and of a connection on it: the trivial connection is supplied
by inertia, since inertial r(t)'s are straight, i.e. inertial worldlines are
geodesics of spacetime and the non-trivial connection is supplied by
gravity. However, there is no spacetime metric even in Newton-Cartan
theory - Einstein's general relativity however adds a spacetime metric,
and that provides its essential advantage . Furthermore Einstein theory
requires 4 dimensional general coordinate invariance as opposed to just 3
dimensional general coordinate invariance. Our approach will be using Newtonian results and then a 4 dimensional metric interpretation of Newtonian results plus the requirement of smooth
matching of metrics, like the cosmological solution inside a ball of dust
to an unknown static spherically symmetric metric outside. In order to
construct the smooth matching of these two space times, it is essential to
transform the inside metric to a coordinate system such that the two
metrics, inside and outside smoothly match. Therefore, we crucially use
the concept of four dimensional general coordinate invariance, but not the
full Einsteins equations, so we reproduce GR results, not Einstein Cartan
results.

\medskip
By matching boundary conditions for a collapsing dust ball and its exterior vacuum, i.e. an interior FLRW metric to a Schwarzchild exterior, we can specify elements of one of the metrics in terms of the other, and so we needn't fully solve the Einstein's equation (EE) for both metrics. So for example we can find $g_{rr}$ of the Schwarzschild metric by knowing its $g_{00}$, and knowing $R(t)$ of the FRWL metric, so that the spatial curvature aspect of the Schwarzschild metric is specified by the non-spatial-aspects of the interior (FLRW metric) and exterior (Schwarzschild metric). As it turns out, we can use a combination of GR and Newtonian theory arguments to produce $g_{00}$ of Schwarzschild without using the EE, and a combination of GR and Newtonian-cosmology theory arguments which produces the Einstein equation governing $R(t)$ without actually needing the EE itself, and so we do not actually need the EE themselves to specify both metrics completely. 

\medskip
As mentioned earlier (Also in the abstract, or in the introduction), the
paper arrives at results for the Schwarzschild metric (finding $g_{rr}$) by
using Boundary Conditions with a solution for a collapsing ball of dust, which is also a
cosmological solution, and with this method we also arrive at an
interpretation of the "K" in the FRLW metric for cosmology. Therefore we
will be jumping around in the paper between results from both Schwarzschild
and a collapsing dust-ball or cosmology.Here, we shall also present not just
the new results but also known ways of arriving at some of the results we
will be using along the way.

\medskip
 we can try to make this list of important points of various sections and indicate which is a new result and which is not, and that it follows the order of the sections in the paper:\\

   1. We shall show various ways (some standard, some only presented here, some
   more rigorous and some very heuristic) in which one can arrive at $g_{00}$ of
   Schwarzschild without using the Einstein's equation in sec 4.\\
   2. We shall show a standard means of arriving at the Friedman equations for
   cosmology, to use it later on in the paper.\\
   3. We match BCs between interior and exterior of a dust ball and find a relation between the coordinates for the two relevant solutions which will serve us in finding our desired result.\\
   4. Using the results from the section 2 we shall derive and present
   a new result, finding $g_{rr}$ of Schwarzschild without use of the Einstein's equation.\\
   5. Using the results from the sections 2 \& 5, we shall find a
   novel spatial-curvature interpretation of the K in the Newtonian-derived
   Friedman equation.\\
   6. In conclusion, we shall comment on our over-all work and results.\\

\section{Formulation of the problem}

 In this paper ( also see \cite{12},\cite{13} ), we shall develop an approach which will go beyond the simple derivation of the $0-0$ component of the Schwarzschild metric  and give a derivation to obtain the $g_{\bar{r}\bar{r}}$ component under some assumptions that we shall specify properly and shall avoid previous criticism\cite{2}.
 
 \begin{figure}[ht!]
\begin{center}
\includegraphics[width=175mm]{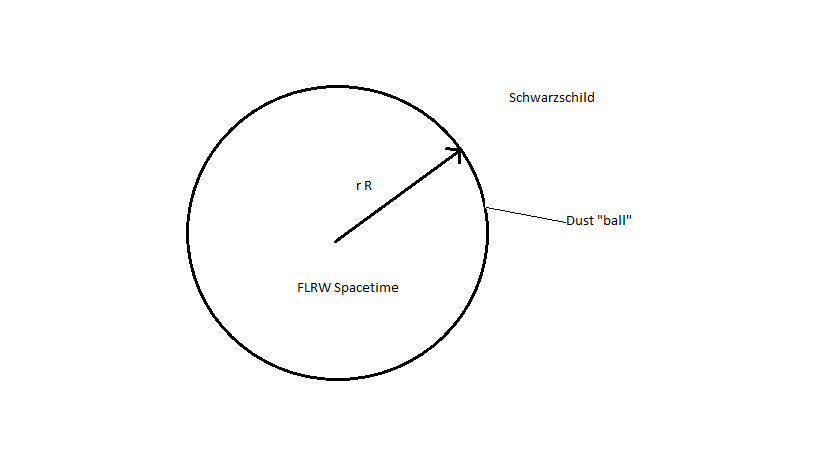}
\caption{The dust "ball" with outside Schwarzschild metric and inside FLRW metric }
\end{center}
\end{figure}
 
 \medskip
Our methods to discover the geometrical meaning of Newtonian cosmology and
the full Schwarzschild metric without the usage of Einstein's equation, it is indeed possible to obtain the Friedmann equations for the expansion
factor $R(t)$ in the simple investigation process of this paper. We are going to deal with a cosmological "ball " of matter. Outside this spherically symmetric ball we will assume there is a spherically symmetric static space time and this ball is free falling. Here, we have used the analogy of Newtonian gravity or electromagnetism where outside the matter or conducting sphere the field is static. We also have assumed the the expansion of the matter dust balls along-with a spherical symmetry in  consistency with the global isotropy. As is well known, any static metric can be brought to the "standard" form. Here-forth, we shall be labeling coordinates of the outside space using  bars, so we define  $\bar{t}$ and $\bar{r}$ for the time and radial coordinates outside the spherical ball. So, the outside metric reads as follows:
 
\begin{eqnarray}
\label{externalmetric}
ds^2 = - B(\bar{r})d \bar{t}^2 + A(\bar{r})d \bar{r}^2 + \bar{r}^2\left(d\theta^2 + \sin^2 \theta d \phi^2\right).
\end{eqnarray}

By the requirement that Newton's law for gravity holds in the limit, we can obtain  an idea of what
 $g_{00} = - B$ 
should be. This temporal  metric component can be obtained by the requirement that the geodesic of slowly moving particles match the Newtonian equation. For  $g_{\bar{r}\bar{r}} = A(\bar{r})$ is harder to investigate from any such assumptions.

\medskip
 As it is known that Newtonian cosmology \cite{4} can be developed and a homogeneous dust system can be studied and by applying the Newton’s laws. We can obtain that each dust particle expands or contracts according to a single expansion factor $R(t)$, obeying
\begin{eqnarray}
\left( \frac{\dot{R}}{R} \right)^2 = \frac{8 \pi G}{3} \rho - \frac{k}{R^2},
\end{eqnarray}

where, $k$ being an integration constant. Points to be noted that this equation came out from the second order differential equation of R.

\medskip
The radius for a given particle should be proportional to $R(t)$. By using the continuity equation and the Euler motion equation, we can obtain that the density $\rho \sim \frac{\rho_0}{R^3}$ and if we require boundary conditions such as $R(0) = 1, \dot{R}(0) = 0 , \rho = \rho(0)$ at $t=0$, we get
\begin{eqnarray}
\label{the value of k}
k = \frac{8 \pi G}{3} \rho(0) > 0 \quad \textsl{and} \quad \dot{R}(t)^2 = k \left[\frac{1}{R} - 1 \right].
\end{eqnarray}

In this section, we shall not focus on or derive or motivate a geometric interpretation of the constant of integration $k$. In contrast, a follow up section will see that, even with out using Einstein's Equation, $k$ can be given the interpretation of spatial curvature in FLRW space. Then matching to an external space can be used to determine the outside space by demanding continuity in some coordinate system from the inside to the outside of the "ball". Of course, the fact that one can give a geometrical interpretation to the constant $k$ without using Einstein's Equation is by itself very interesting and intriguing.

 Here we only are going to assume that the metric inside the ball is of the form
\begin{eqnarray}
\label{dt2}
ds^2 = -dt^2 + g_{ij} dx^i dx^j \ \forall i,j=1(1)3
\end{eqnarray}

and that the dust particles are co-moving (The metric depends on the coordinate frame of the observer or observer is freely falling ) so that $dx^i=0$ which are geodesics of the space time\cite{3}. Furthermore, just from homogeneity and isotropy, we can determine that 
\begin{eqnarray}
\label{gij}
g_{ij} dx^i dx^j = R^2(t)\left(\frac{dr^2}{1-\kappa r^2} + r^2\left(d\theta^2 + \sin^2\theta d\phi\right)\right)
\end{eqnarray}
where the radial and time coordinates are $r$ and $t$ respectively. Notice that apriori we do not know if $k$ and $\kappa$ are related.

\section{Newtonian Cosmology}
In this section, we shall reproduce some basic equations of Cosmology from basic assumptions of Newtonian Physics. In a homogeneous and isotropic universe, we can always assume from the cosmological principle, a spherical region of radius $R$ around an arbitrary point, whose matter density $\rho$ within is homogeneous and isotropic. The surrounding matter cannot have any influence on its dynamics, as this would violate isotropy. In this general assumptions, the size of the sphere is completely arbitrary.
The equation of motion for a test mass $ m $ located at the boundary of such a sphere shall be described in terms of a homogeneous positive parameter $R(t)$, where the coordinate of each particle expands according to $a(t)=constant \cdot R(t)$, this constant depends on the particular particle. Along with the assumption of the motion of the particles as follows (see \cite{4},\cite{14},\cite{15}). The velocity is divided into temporal function and the radial vector whereas density and pressure are spatial independent.We can deduce the claim of velocity as a function of radial part and function of time, we proceed in a following way :

\medskip
\subsection{\bf{Acceleration Equation and time evolution of the dust density}}

Consider a particle of mass m is residing outside the sphere. So,
the attraction due to gravity should hold the equation

\begin{equation}
m\ddot{r}=\frac{GMm}{r^{2}}
\end{equation}

\begin{equation}
\Rightarrow\ddot{r}=\frac{GM}{r^{2}}
\end{equation}

\begin{equation}
\Rightarrow\ddot{R}=\frac{4}{3}\pi GR(t)\rho(t)
\end{equation}

Now, Since the mass of the Volume is constant 

Therefore, 
\begin{equation}
M=\rho V=\rho(t_{0})V(t_{0})
\end{equation}

so that 
\begin{equation}
\rho(t)=\rho(t_{0})\frac{R^{3}(t_{0})}{R^{3}(t)}
\end{equation}

So, our acceleration term becomes 

\begin{equation}
\ddot{R}=\frac{4}{3}\pi G(t_{0})\rho(t_{0})\frac{R^{3}(t_{0})}{R^{2}(t)}
\end{equation}

All our derivation were based on assumption that fluid is Pressureless. For added pressure Fluid equation derivation, please consult the appendix.

\medskip
\subsection{\bf{Hubble's law}}

If we denote the coordinate of the co-moving observer as unprimed whereas
the coordinate of a moving object with respect to the observer as
primed. Then, the relative velocity of the object with respect to
the observer can be written as $\overrightarrow{v^{\prime}}=\overrightarrow{v^{\prime}}(\overrightarrow{r^{\prime}},t)$. Now, if we take our origin to a general point and if the relative
distance between the observer and the object is $\overrightarrow{s}$
then the velocity vector and displacement vector becomes 
\\
$\overrightarrow{v^{\prime}}(\overrightarrow{r}-\overrightarrow{s})$
and $\overrightarrow{r^{\prime}}=\overrightarrow{r}-\overrightarrow{s}$
using Galilean Transformation.
\\

Therefore 

\begin{equation}
\overrightarrow{v^{\prime}}(\overrightarrow{r}-\overrightarrow{s})=\overrightarrow{v}\overrightarrow{(r)}-\overrightarrow{v}\overrightarrow{(s)}
\end{equation}

This implies the transformation is indeed linear. This leads us to
conceive $\overrightarrow{v}\overrightarrow{(r)}=f\overrightarrow{r}$. Where, $f$is a function independent of $\overrightarrow{r}$. It must be dependent solely on t, so that  $f=f(t)$.

So we can write them in following way :

 \begin{equation}
\overrightarrow{v}=f(t)\overrightarrow{r},
\end{equation}

\begin{equation}
\rho=\rho(t),
\end{equation}

\begin{equation}
P=P(t)=0,
\end{equation}

here we are discussing the situation with $P=0$.
\\

The first of the above mentioned three equation after integration
gives rise to 

\begin{equation}
\overrightarrow{r}=R(t)\overrightarrow{r_{0}}
\end{equation}

where, $R(t)$ satisfies 

\begin{equation}
\frac{1}{R}\frac{dR}{dt}=f(t)\;,R(t=0)=1
\end{equation}
\medskip
This equation is a reminder of Hubble's law with the Hubble's parameter $H(t) = f(t)$. $R(t)$ is nothing but a scale factor reminder of isotropic expansion of the universe.

\medskip
The proper distance at any particular time therefore is Proportional
to the scale factor at that time. So that,

\begin{equation}
\frac{D(t)}{D(t_{0})}=\frac{R(t)}{R(t_{0})}
\end{equation}
\medskip

\subsection{\bf{Friedmann Equation}}

Now by applying Euler's equation of motions , we have 

\begin{equation}
\frac{D\overrightarrow{v}}{Dt}-\overrightarrow{F}=0=\overrightarrow{r}[\frac{df}{dt}+f^{2}]-\overrightarrow{F},
\end{equation}

where the Hydrodynamic operator is $\frac{D}{Dt}=\frac{\partial}{\partial t}+(\vec{v.}\nabla)$,

 taking divergence on both sides 

\begin{equation}
3[\frac{df}{dt}+f^{2}]=\nabla \overrightarrow{F}
\end{equation}

where, $\overrightarrow{F}$ is the body force per unit mass.

Poisson's equation in its usual form:

\begin{equation}
\nabla.\overrightarrow{F}=-4\pi G\rho
\end{equation}

So, we have now then

\begin{equation}
3[\frac{df}{dt}+f^{2}]=-4\pi G\rho
\end{equation}

Putting the value of $f(t)$ back yields a similar equation for the universal expansion factor $R(t)$

\begin{equation}
\label{eq:newton1}
\ddot{R}=-\frac{G}{R^2} \left(\frac{4\pi}{3}R^3 \rho \right)= - \frac{4\pi G}{3} R\rho.
\end{equation}

Basically, this corresponds to Friedmann's second equation without a cosmological constant $\Lambda$ and zero pressure.
Note: as the linear dimensions has been scaled by $R(t)$, all co-moving volumes should be  scaled by $R(t)^3$ and therefore a $1/R^{3}$ dependency is for the density, which dilutes the matter as the sphere expands.

For deriving a second equation, we first consider mass conservation within co-moving sphere\cite{10},
\begin{equation}
\label{eq:energyconservation1}
\frac{d}{dt}\left(\frac{4\pi}{3}  R^{3}\rho  \right) = 0,
\end{equation}
where the internal mass $M$ inside the sphere should be conserved. By performing the derivative and simplifying one $R$, the equation gets
\begin{equation}
2R\dot{R}\rho + R\dot{R}\rho + R^2 \dot{\rho} = 0.
\end{equation}
The second term can be eliminated by (\ref{eq:newton1}) and after restoring derivatives the equation
\begin{equation}
\frac{d(\dot{R}^2)}{dt} = \frac{8\pi G}{3} \frac{d(\rho R^2)}{dt}
\end{equation}
is obtained. Integration on both sides gives
\begin{equation}
\dot{R}^2 = \frac{8\pi G}{3} \rho R^2 - \tilde{k},
\end{equation}
and rewriting the arbitrary integration constant $\tilde{k}$ in a way to match the units $\tilde{k} \longrightarrow kc^2$ yields finally an equation, which corresponds to Friedmann's first equation  \cite{10} :  
\begin{equation}
\label{eq:fridman1}
\left(\frac{\dot{R}}{R}\right)^{2} = \frac{8\pi G}{3} \rho - k \left(\frac{c}{R}\right)^2.
\end{equation}

If we think, it is very much like a conserved energy of a mechanical system and
indeed the sign of k determines whether R will expand to infinity, or expand, achieve a
maximum and then re collapse. This can be compared with a mechanical system in a
following explanatory way:

\medskip
-For $k>0$, we have a bound system, a bound system that has an associated negative energy\\
-For $k<0$, represents a case where the expansion factor reaches infinity like that of a mechanical system    with positive energy.\\
-For$k=0$, produces the case where R barely makes it to infinity and corresponds to a system with zero energy, analogous to the very similar problem of the critical escape trajectory that just makes in order to escape the earth.

\medskip
A possible relation between $k$ and the $\kappa$ that appears in (\ref{gij}),can be  achieved without assuming Einstein's equations. The FLRW and the appearance of $R(t)$ in (\ref{gij}) can be justified on the ground of  purely symmetry and geometrically, independent of any specific dynamics, general relativity or anything such.

As we pointed out before, although we shall not invoke Einstein's equations, we shall make use of the much simpler geodesic equation. As we mentioned, it is a rather simple matter to show that "co-moving observers" of the metric (\ref{dt2}), (\ref{gij}) follows those trajectories where the spatial coordinates 
(\ref{dt2}),(\ref{gij}) are constants or indeed geodesics. Since, $x^{i}=const$ is a geodesic in FLRW spacetime as FLRW metric yields $\Gamma_{tt}^{\mu}=0$ \cite{7}. Physical distances between these  "co-moving observers" scale as $R$, so the $R$ in the Newtonian cosmology coincides with the $R$ that appears in (\ref{gij}).

Now, the relation between $k$ and $\kappa$, this will require a bit more elaboration (without using Einstein's Equation). We can now describe cosmic large scale dynamics by a single expansion parameter $R(t)$, as long the cosmic principle holds, and the predictions corresponds to general relativity. Note that we are using $c=1$.

\medskip
Finally, we note that in Newtonian cosmology, the following equations hold invariant under a transformation to an accelerated  coordinate frame
:

\begin{equation}
\ddot{X}_{i}+\frac{\partial}{\partial X_{i}}\phi(X_{j},t)=0=\ddot{X}'_{i}+\frac{\partial}{\partial X'_{i}}\phi(X'_{j},t)
\end{equation}
 where in the transformation we introduce an arbitrary uniform acceleration in the following way:

\begin{equation}
\ddot{X}_{i}'(t)-\ddot{X_{i}}(t)=g_{i}(t)
\end{equation}

\begin{equation}
\phi'(X_{j}',t)=\phi[X_{i}(x_{j}'),t]-g_{i}(t)X_{k}(X_{j}')+h(t)
\end{equation}

$\forall i,j,k=1,2,3.$
 
\medskip

The gravitational potential can be found to be of the form 
\begin{equation}
\phi=\frac{2\pi G\rho}{3}X_{i}X_{i}
\end{equation}

 which doesn't single out a special point in the universe, since, as pointed out  in \cite{18}, the transformations (30), (31) imply all points are on an equal footings, so the Potential is good for any arbitrary origin.
 
 \medskip
 
 Notice that the $ X_{i} $ coordinates are related to the co-moving coordinates ( $ x_{i} $ ) by means of the expansion  factor  through the relation $X_{i}= R(t)x_{i}$. So for the isotropic and homogeneous cosmology, where  $x_{i}= constant$, we have that a transformation $x'_{i} = x_{i} + a_{i}$, where $a_{i} = constant$ is in terms of the $X_{i}$ coordinates a transformation to another accelerated frame. Under such transformation, the extra induced terms in (32) can be eliminated by the transformation (31), so, this indeed provides a proof that
the Potential is good for any arbitrary origin.
\section{To find $g_{00}$}

In this section, in order not to make the notation to heavy, we will for the moment denote the coordinates associated with the static space time simply t, r, $\theta$ and $\phi$. In the next section, we go back to the bar notation when it refers to an external space and unbarred coordinates for the internal cosmological space time. In this section, we are not going to derive anything new, rather in short going to show the connection between the temporal component of Scwarzschild metric and Newtonian Gravity. We are going to reproduce the temporal component of the metric by  explicitly  metric perturbation method using traditional non relativistic and static assumptions and we shall leave as a problem to find out $g_{00}$ by Gravitational Potential with sufficient hints and references . Both of them, for obvious reason, shall lead us to the same result.

\medskip
We shall derive the $g_{tt}$ element of the metric from metric perturbation method while same result can be achieved in the non relativistic limit. We shall take into account of free fall particles having geodesic worldlines, whether the spacetime is warped in the presence of matter, and free fall worldlines are geodesics of the warped spacetime, and so we can use the geodesic equation to find an equation of the free fall worldline, which gives us the 4-acceleration (the geodesic equation says this acceleration is 0, i.e. the worldline curvature = 0) and thus we can relate it to the Newtonian gravity equation for the acceleration of a free particle (in Newtonian gravity the interpretation is that a free fall particle has acceleration $\Gamma$, rather than $\Gamma$ being a correction factor in the equation with acceleration = 0). All this can be stated in a bit different way, in the Newtonian regime the geodesic equation reduces to $\frac{d^{2}r}{dt^2} = \Gamma ^{r}_{tt}$, which in Newtonian gravity is the acceleration equation 
the observer standing on Earth watching the free fall particle sees it has acceleration, but it is the observer's acceleration, the free fall particle has zero acceleration: the Earth-observer's world-line is curved, and that curvature is his acceleration, and it is exactly the amount of the acceleration between him and the free fall particle, whoever is considered to be accelerating. The free fall particle's equation is the geodesic equation, saying that its world-line curvature is zero, and it vanishes due to the correction factor in the geodesic equation, the Christoffel connection, so that amount is exactly the acceleration as seen from the Earth-stationary viewpoint.
\\

\subsection{\bf{ Metric perturbation}}

\medskip
We, now, allow our usage of the geodesic equation, because this subject does not involve a high level of complexity as compared to the development of the Einstein's equations, and it can be derived from the  from the principle of least action of the particle trajectory, the action being the proper time along the trajectory of the particle in a certain given metric. Starting from the interval
\medskip
\begin{eqnarray}
 ds^{2}=g_{\mu\nu}dx^{\mu}dx^{\nu} = -d\tau^2,
\end{eqnarray}
\medskip
we integrate over $d\tau$ and by using a monotonic variation parameter $\sigma$ on the particle trajectory
\begin{eqnarray}
\mathcal{\tau}=\int\mathbf{d}\sigma\sqrt{-g_{\mu\nu}\frac{dx^{\mu}}{d\sigma}\frac{dx^{\nu}}{d\sigma}},
\end{eqnarray}
from here applying the principle of least action we find the geodesic equation
\begin{eqnarray}
 \frac{d^{2}x^{\mu}}{d\sigma^{2}} + \Gamma_{\alpha\beta}^{\mu}\frac{dx^{\alpha}}{d\sigma}\frac{dx^{\beta}}{d\sigma} = 0,
\end{eqnarray}
with $\Gamma_{\alpha\beta}^{\mu}$ known as the Christoffel symbol (or connection)
\begin{eqnarray}
 \Gamma_{\alpha\beta}^{\mu}=\frac{1}{2}g^{\mu\rho}\left(\frac{\partial g_{\rho\beta}}{\partial x^{\alpha}}+\frac{\partial g_{\alpha\rho}}{\partial x^{\beta}}-\frac{\partial g_{\alpha\beta}}{\partial x^{\rho}}\right).
\end{eqnarray}
Remember that $\Gamma_{\alpha\beta}^{\mu}$ is not a tensor, but the sum of the two terms in the geodesic equation is a vector.

\medskip
The perturbative method starts with taking two assumptions :
\begin{enumerate}
\item  The metric is static with a symmetric perturbation 
\begin{eqnarray}
g_{\mu\nu}=\eta_{\mu\nu}+h_{\mu\nu}\;;\;\left|h_{\mu\nu}\right|\ll1 \quad \textsl{and} \quad g_{\mu\nu}\approx\eta_{\mu\nu}\;;\; g^{\mu\nu}\approx\eta^{\mu\nu},
\end{eqnarray}
\item Non-relativistic limit
\begin{eqnarray}
v^{i}=\frac{dx^{i}}{dt}\ll1\rightarrow\frac{dx^{i}}{d\tau}\ll\frac{dt}{d\tau}\sim1,
\end{eqnarray}
\end{enumerate}
We now return to the geodesic equation
\begin{eqnarray}
 \frac{d^{2}x^{\mu}}{d\tau^{2}}=-\Gamma_{\alpha\beta}^{\mu}\frac{dx^{\alpha}}{d\tau}\frac{dx^{\beta}}{d\tau}=-\Gamma_{tt}^{\mu}\frac{dx^{t}}{d\tau}\frac{dx^{t}}{d\tau}-\Gamma_{ij}^{\mu}\frac{dx^{i}}{d\tau}\frac{dx^{j}}{d\tau}\approx-\Gamma_{tt}^{\mu},
\end{eqnarray}
writing the equations for the spatial part
\begin{eqnarray}
 \Gamma_{tt}^{i}=-\frac{1}{2}\eta^{ij}\left(\frac{\partial g_{tt}}{\partial x^{j}}\right)=\frac{1}{2}\delta^{ij}\left(\frac{\partial g_{tt}}{\partial x^{j}}\right)
 \frac{d^{2}x^{i}}{dt^{2}}=\frac{1}{2}\left(\frac{\partial g_{tt}}{\partial x^{i}}\right).
\end{eqnarray}
Now, by demanding the asymptotic limit:  as ${r\rightarrow\infty}$ , $g_{tt}\rightarrow\-1$, 

As we already know from Newtonian force law of universal gravity,

\begin{eqnarray}
\frac{d^2 r}{dt^2} = - \nabla \phi \quad \textsl{where} \quad \phi = -\frac{GM}{r},
\end{eqnarray}

So, finally combining equation (39) and (40), and comparing to (41) with the usage of the Boundary Condition at infinity, we find that we have the $0-0$ component of metric of the following form:

\begin{eqnarray}
 g_{tt}=-\left(1-\frac{2GM}{r}\right)=-\left(1+2\phi\right).
\end{eqnarray}

\subsection{\bf{Problem}}
Using that the gravitational acceleration of a radically falling object is given by $\ddot{r}=-\frac{GM}{r^2}$, we obtain that $r\propto t^{2/3}$. If we want to re-express this as an " inertial " motion, we must use a new time $t'$ such that $r\propto t'$. Now, by considering $dt'^2$ and expressing $dt'^{2}$ as $f(r)dt^2$ and then we can reach to an expression that can be the motivation of finding Scwarzschild temporal metric. For more see \cite{17}.

\medskip
-it should tally with the result of previous subsection.

\section{To find $g_{\bar{r}\bar{r}}$}
This section is devoted to finding the $r-r$ component of metric by the matching of boundary condition at the surface of the Spherical Ball. Since, from exactly Newtonian Gravity standpoint of view , this is quite impossible to find $g_{rr}$ component of Scwarzschild metric, so this section plays a very important role in adding value of this current work. The  $g_{\bar{r}\bar{r}}$ component of the metric in the outside static region in our case has been the "elusive component". This component has not been calculated using matching to a Newtonian cosmology previously and here we shall show that this is indeed possible. In this section we are going to find out  $g_{\bar{r}\bar{r}}$ using the assumption that we have a co-moving observer satisfying $r = constant$. We will also assume that inside and at the boundary of the dust ball the radius evolves as $\bar{r}= r R(t)$, where $R(t)$ is determined by (\ref{eq:fridman1}). We also assume that for $\bar{r} > r R(t)$ the motion is on a radial geodesic and the metric is of the form
\begin{eqnarray}
ds^2 &=& - \left( 1 - \frac{2GM}{\bar{r}}\right) d\bar{t}^2 + A(\bar{r}) d\bar{r}^2 + \bar{r}^2 d \Omega^2 \\
d \Omega^2 &=& d \theta^2 + \sin^2 \theta d \phi^2 ,
\end{eqnarray}
we labeled the time as $\bar{t}$ because it may not be the same coordinate as the $t$ in the dust ball and the  $g_{tt}$ component was found in the previous section. Notice that we assume that the external metric is time independent.

\medskip

A radially falling geodesic, meaning that $\theta = const$ and $\phi = const$, is fully described by the conservation of energy that results from that the metric outside is assumed to be static.
The geodesics are derived from the action 

\begin{eqnarray}
S = \int d\sigma \sqrt{-\frac{d x^\mu}{d \sigma} \frac{d x^\nu}{d \sigma} g_{\mu \nu}(x)}.
\end{eqnarray}
\medskip
The equation with respect to $\bar{t}$ is

\begin{equation}
\frac{d}{d\sigma}\left( \frac{\partial L}{\partial \dot{\bar{t}}}\right) = 0 \quad \textsl{where} \quad \dot{\bar{t}} = \frac{d \bar{t}}{d \sigma}.
\end{equation}

This gives us
\begin{equation}
\gamma = \frac{\partial L}{\partial \dot{\bar{t}}} = \left( 1 - \frac{2GM}{\bar{r}} \right) \frac{d \bar{t}}{d \tau},
\end{equation}
where $\gamma$ is constant and $d\tau$ is the proper time.\\
Notice that since the spatial coordinates in the space $d x^i = 0$ we get 
\begin{equation}
d\tau = d t ,
\end{equation}
giving us
\begin{eqnarray}
d \tau^2 &=&  \left( 1 - \frac{2GM}{\bar{r}} \right) d \bar{t}^2 - A(\bar{r})d\bar{r}^2\\
\left( \frac{d \tau}{d t} \right)^2 = 1 &=& \left( 1 - \frac{2GM}{\bar{r}} \right) \left( \frac{d \bar{t}}{dt} \right)^2 - A(\bar{r}) \left( \frac{d\bar{r}}{dt} \right)^2 .
\end{eqnarray}
Using previous equations, we obtain
\begin{eqnarray}
\left( 1- \frac{2GM}{\bar{r}}\right)^{-1}\gamma^2 - A(\bar{r}) \left( \frac{d\bar{r}}{dt} \right)^2 = 1 .
\end{eqnarray}

\medskip
As we have seen, the consistency of the matching of the two spaces requires
 $\bar{r} = R(t) r$, furthermore, we assume that even the boundary of the dust shell
 free falls according to a co-moving observer, which means that the FLRW coordinate
 $r= constant$ and this allow then to solve for  $A(\bar{r})$, 
\begin{eqnarray}
A(\bar{r}) = - \left( 1 - \frac{\gamma^2}{\left( 1 - \frac{2GM}{\bar{r}} \right)}\right)\frac{1}{r^2 \left( \frac{d R}{dt}\right)^2}.
\end{eqnarray}

This easily yields
\begin{eqnarray}
A(\bar{r}) = -\left( 1 - \frac{\gamma^2}{\left( 1 - \frac{2GM}{\bar{r}} \right)}\right)\frac{1}{r^2 k \left( \frac{1}{R} -1 \right)}
\end{eqnarray}
by simplifying this and expressing in terms of $\bar{r}$ we get
\begin{eqnarray}
A(\bar{r}) = \frac{1}{r^2}\frac{\gamma^2 - 1 + \frac{2GM}{\bar{r}}}{1 - \frac{2GM}{\bar{r}}}\frac{1}{k \left( -1 + \frac{r}{\bar{r}}\right)}.
\end{eqnarray}

If we take the limit $\bar{r} \rightarrow \infty $, we see that 
$A(\bar{r}) \rightarrow -(\gamma^2 - 1)/k r^2 $. Asymptotic flatness would require 
$A(\bar{r}) \rightarrow 1$. 

\medskip
The metric component $A(r)$ is free of singularities (real singularities, not coordinate singularities) and preserves its signature (one time and three spaces)and is asymptotically flat only if
\begin{eqnarray}
\label{A}
\gamma^2 - 1 = - k r^2 \quad \textsl{and} \quad \frac{2GM}{\bar{r}} = \frac{r^2 k}{R} = \frac{k r^3}{\bar{r}} \Rightarrow k = \frac{2GM}{r^3}.
\end{eqnarray}
As mentioned earlier, $r$ is the comoving coordinate and hence constant, whereas $R$ is variable.

\medskip
Notice that above the condition,(\ref{A}) $ k = \frac{2GM}{r^3}$ , when combined with the value of k, as given by
(\ref{the value of k}), $k = \frac{8 \pi G}{3} \rho(0)$ yields us the value of M as
\begin{eqnarray}
M=\frac{4}{3}\pi\rho_{0}r^{3}.
\end{eqnarray}

Surprisingly, it looks like the $Mass= density \times Vol. of E^{3} ball. $ But, we got this relation in consequences of our previous derivation and the flat space volume of the ball has not been used anywhere for deriving this. 

\medskip
Finally, all of this gives us
\begin{eqnarray}
A(\bar{r}) = \frac{1}{1 - \frac{2GM}{\bar{r}}},
\end{eqnarray}
reproducing the Schwarzschild spacetime.

\section{To find the geometric interpretation of the Newtonian Cosmology}
The main aim of this section is to show that FRLW $\kappa$ is equal to Newtonian cosmology $k$.
\medskip
In an embedding n dimensional Euclidean Spacetime ($R^{n}$) with metric
\[
ds^2 =dx^{i}dx_{i} , \forall i=1(1)n
\]
we define a submanifold of one sheet ($x_i$) satisfying the constraint
\begin{eqnarray}
x^{i}x_{i}=1/ \kappa 
\end{eqnarray}
with $\kappa >0$, it is said to form a ${ S^{n-1}}$  manifold.

\medskip
So, we will take our cosmological solution to have positive spatial curvature $\kappa>0$ 
therefore topologically spatial slices shall have the topology of $S^{n-1}$.
 
\medskip
To obtain a space time, that is the line element in n-d  Spacetime, we add time and a scale factor that multiplies the $n-1$-sphere, 
we obtain,

\begin{equation}
ds^{2}=-dt^{2}+R^{2}(t)dS_{n-1}^{2} .
\end{equation}

\medskip
In an embedding four dimensional Euclidean Space with metric (n=4)
\begin{equation}
ds^2 = dx^{2}+dy^{2}+dz^{2}+dw^{2},
\end{equation}

\medskip
we define a 3-sphere if the sets of points ($x,y,z,w$) satisfy
the constraint
\begin{equation}
x^{2}+y^{2}+z^{2}+w^{2}=1/ \kappa 
\end{equation}
with $ \kappa >0$
then it is said to form a three - sphere $S^{3}$.

\medskip
That is, we will take our cosmological solution to have positive spatial curvature $\kappa>0$ therefore topologically 
spatial slices have the topology of $S^{3}$.
Solving for $w$ from the above constraint and inserting into the expression for $dl^2$  
defining $r^2 = x^{2}+y^{2}+z^{2}$,  $x=rsin\theta cos \phi$
and $y=rsin\theta sin\phi$,$z=rcos\theta$, we obtain the metric of the $3$-sphere,
\[
ds_{3}^{2}=[\frac{dr^{2}}{1-\kappa r^{2}}+r^{2}d\Omega^{2}]
\]
with 
\begin{equation}
d\Omega^{2}=d\theta^{2}+sin^{2}\theta d\phi^{2}
\end{equation}

to obtain the physical space time, that is the line element in FLRW Space, we add time and a scale factor that multiplies the $3$-sphere, 
we obtain,
\[
ds^{2}=-dt^{2}+R^{2}(t)ds_{3}^{2} .
\]
So our infinitesimal line element in
FRW space becomes, 
\begin{eqnarray}
ds^{2}=-dt^{2}+R(t)^{2}\left(\frac{dr^{2}}{1-\kappa r^{2}}+r^{2}d\Omega^{2}\right).
\end{eqnarray}
\medskip
A very basic and elementary relation we obtain by matching the space time (\ref{externalmetric})
with (\ref{dt2} - \ref{gij}),  along with considering the length of an equatorial circle
($sin\theta =1$)as $\phi$ runs from $0$ to $2\pi$. The inside observer (in the dust ball), just below the surface of matching will measure a length $2\pi r R$, while The outside observer (in the static space), just above the surface of matching will measure a length $2\pi \bar{r}$, since these two measurements refer to the same physical length, in order for the complete geometry to be well defined, we obtain
\begin{eqnarray}
\label{bar}
\bar{r} = r R(t).
\end{eqnarray}

Thus we have established a relationship between the outside radial coordinate and the expansion factor.

\medskip
We assume again that that we have a co-moving observer which satisfies $r = const$. Independently of that,
in the FLRW space we can use everywhere (not just at the boundary) the barred radius  $\bar{r}=R(t)r$, which means $r=\frac{\bar{r}}{R(t)}$.

\medskip
This yields

\begin{eqnarray}
dr=\frac{d\bar{r}}{R(t)}-\frac{\dot{R\bar{r}}}{R^{2}}dt.
\end{eqnarray}
Putting this in $(69)$, we've
\begin{eqnarray}
ds^{2}=\left(-1+\frac{\dot{R}^{2}\bar{r}}{(1-\kappa r^{2})R^{2}}\right)dt^{2}+\frac{d\bar{r}^{2}}{1-\kappa r^{2}}-\frac{2\dot{R}\bar{r}}{(1-\kappa r^{2})R}dtd\bar{r}.
\end{eqnarray}
Now, we make a transformation $t=t(\bar{t},\bar{r})$, so infinitesimal change in time 
\begin{eqnarray}
dt=\frac{\partial t}{\partial\bar{r}}d\bar{r}+\frac{\partial t}{\partial\bar{t}}d\bar{t},
\end{eqnarray}
squaring, 
\begin{eqnarray}
dt^{2}=\left(\frac{\partial t}{\partial\bar{r}}\right)^{2}d\bar{r}^{2}+\left(\frac{\partial t}{\partial\bar{t}}\right)^{2}d\bar{t}^{2}+2\left(\frac{\partial t}{\partial\bar{r}}\right)\left(\frac{\partial t}{\partial\bar{t}}\right)d\bar{t}d\bar{r}.
\end{eqnarray}

\medskip
Under these circumstances, the infinitesimal line element becomes

\begin{equation}
ds^{2}=\frac{d\bar{r}^{2}}{1-\kappa r^{2}}-\frac{2\dot{R}\bar{r}}{\left(1-\kappa r^{2}\right)R}d\bar{r}\left(\frac{\partial t}{\partial\bar{r}}d\bar{r}+\frac{\partial t}{\partial\bar{t}}d\bar{t}\right)\\
+\left(-1+\frac{\dot{R}^{2}\bar{r}}{(1-\kappa r^{2})R^{2}}\right)\left(\left(\frac{\partial t}{\partial\bar{r}}\right)^{2}d\bar{r}^{2}+\left(\frac{\partial t}{\partial\bar{t}}\right)^{2}d\bar{t}^{2}+2\left(\frac{\partial t}{\partial\bar{r}}\right)\left(\frac{\partial t}{\partial\bar{t}}\right)d\bar{t}d\bar{r}\right)
\end{equation}

\begin{eqnarray}
\nonumber=\left\{ -1+\frac{\dot{R^{2}}\bar{r}}{(1-\kappa r^{2})R^{2}}\right\} (\frac{\partial t}{\partial\bar{t}})^{2}d\bar{t}^{2}+2\frac{2\dot{R}\bar{r}}{(1-\kappa r^{2})R}+\left[-\frac{2\dot{R}\bar{r}}{(1-\kappa r^{2})R}+2\left\{-1+\frac{\dot{R^{2}}\bar{r}}{(1-\kappa r^{2})R^{2}}\right\}(\frac{\partial t}{\partial\bar{r}})\right]\left(\frac{\partial t}{\partial\bar{t}}\right)dtd\bar{r}+\\
\left[\frac{1}{1-\kappa r^{2}}-\frac{2\dot{R}\bar{r}}{(1-\kappa r^{2})R}(\frac{\partial t}{\partial\bar{r}})+
\left\{ -1+\frac{\dot{R^{2}}\bar{r}}{(1-\kappa r^{2})R^{2}}\right\} (\frac{\partial t}{\partial\bar{r}})^{2}\right]d\bar{r}^{2}
\end{eqnarray}

\medskip
Now, we have to eliminate the cross terms, so

\begin{eqnarray}
-\frac{2\dot{R}\bar{r}}{(1-\kappa r^{2})R}d\bar{r\frac{\partial t}{\partial\bar{t}}d\bar{t}} \quad+2\left(-1+\frac{\dot{R}^{2}\bar{r}}{(1-\kappa r^{2})R^{2}}\right)\left(\frac{\partial t}{\partial\bar{r}}\right)\left(\frac{\partial t}{\partial\bar{t}}\right)d\bar{t}d\bar{r}=0 .
\end{eqnarray}
\medskip
This gives after a careful observation 
\medskip
\begin{eqnarray}
\left(\frac{\partial t}{\partial\bar{r}}\right)=\frac{\bar{r}\dot{R}}{(1-\kappa r^{2})R\left(-1+\frac{\dot{R}^{2}\bar{r}}{(1-\kappa r^{2})R^{2}}\right)}
\end{eqnarray}
\medskip

Lets linger on $g_{\bar{r}\bar{r}}$ with this new $(\frac{\partial t}{\partial\bar{r}})$ value and along with some calculations which can be easily done to see
\\
\begin{eqnarray}
g_{\bar{r}\bar{r}}=\frac{1}{1-\kappa r^{2}}+\left(-1+\frac{\dot{R}^{2}\bar{r}}{(1-\kappa r^{2})R^{2}}\right)(\frac{\partial t}{\partial\bar{r}})^{2}-\frac{2\dot{R}\bar{r}}{(1-\kappa r^{2})R}\frac{\partial t}{\partial\bar{r}}
\end{eqnarray}

\begin{eqnarray}
\qquad=\frac{1}{1-\kappa r^{2}}+\left(-1+\frac{\dot{R}^2\bar{r}}{(1-\kappa r^{2})R^{2}}\right)\left(\frac{\dot{R}\bar{\mbox{r}}}{R^{2}(1-\kappa r^{2})\left(-1+\frac{\dot{R}^{2}\bar{r}^{2}}{(1-\kappa r^{2})R^{2}}\right)}\right)^{2}
-\frac{2\dot{R}\bar{r}}{(1-\kappa r^{2})R}\left(\frac{\dot{R\bar{r}}}{R(1-\kappa r^{2})\left(-1+\frac{\dot{R}^{2}\bar{r}^{2}}{(1-\kappa r^{2})R^{2}}\right)}\right)
\end{eqnarray}

\begin{eqnarray}
=\frac{1}{1-\kappa r^{2}-\frac{\dot{R}^{2}\bar{r}^{2}}{R^{2}}}.
\end{eqnarray}

For $\bar{r}=R(t)r$ it yields

\begin{eqnarray}
g_{\bar{r}\bar{r}}=\frac{1}{1-r^{2}(\kappa+\dot{R^{2}})} .
\end{eqnarray}

In the last section we derived also what this metric component should be,  
\begin{eqnarray}
g_{\bar{r}\bar{r}}=\frac{1}{1-\frac{2GM}{\bar{r}}}.
\end{eqnarray}
Since from Newtonian cosmology, we know that 
\begin{eqnarray}
k+\dot{R}^{2}=\frac{8\pi}{3R}G\rho_{0}=\frac{8\pi r^{3}\rho_{0}}{3r^{2}(rR)}.
\end{eqnarray}
Therefore,
\begin{eqnarray}
(k+\dot{R}^2)r^{2}=\frac{8\pi r^{3}\rho_{0}}{3Rr}=\frac{2GM}{\bar{r}}.
\end{eqnarray}
\medskip
recalling that $M=\frac{4}{3}\pi\rho_{0}r^{3}$ and $\bar{r}=rR$ giving us the relation $k=\kappa$ .

\medskip
A final consistency check is obtained now that we have derived that the internal space is a FLRW spacetime, satisfying the standard Friedmann equations, with k being interpreted as the spatial curvature, as we have just shown. Under this condition, we know the matching of this internal 
cosmology to Schwarzschild is consistent, as the analysis of the Oppenheimer- Snyder collapse model shows\cite{6}.
Here, we have gone through this problem in the opposite way, showing that the matching of these two 
spaces imposes severe constraints that allows us to derive Schwarzschild space outside
and determine that the Newtonian constant of integration $k$ has  to be the spatial curvature of the internal FLRW internal space, all of this without using Einstein's equations.

\section{Conclusions}

Using the Einstein Equations one can of course completely determine the unknown functions in the FRLW and Schwarzschild metrics. And using the known Newtonian gravitational limit one can obtain the constant in the Schwarzschild metric. It is well-known that even without the EE one can obtain $g_{00}$ of Schwarzschild (with issues of large fields etc.), and then the constant as above, leaving only $g_{rr}$ unknown. 
Generally it is assumed that one cannot get farther without the Einstein equations, even by using Newtonian gravity in clever ways since spatial curvature is not part of Newtonian gravity.
It turns out however that one can obtain $g_{rr}$ if one combines the situations - an interior region of collapsing dust with exterior being Schwarzschild - and uses the equations of Newtonian cosmology for the interior and matching of boundary conditions between the two solutions
In this paper it has been found that matching a dust ball, whose dynamics is governed by the Newtonian cosmology equations, containing a constant of integration $k$, to an external static space-time, where 
\begin{eqnarray}
ds^2 = - B(\bar{r}) d \bar{t}^2 + A(\bar{r})d\bar{r}^2 + \bar{r}^2d \Omega^2
\end{eqnarray}
where $B = \left( 1 - \frac{2GM}{\bar{r}}\right)$, forces $A$ to have a very special form. Assuming either asymptotic flatness, or absence of signature change, we uniquely obtain
\begin{eqnarray}
A(r) = \frac{1}{\left(1- \frac{2GM}{\bar{r}}\right)}.
\end{eqnarray}
Finally the same matching to the internal FLRW space
\begin{eqnarray}
ds^2 = -dt^2 + R^2(t)\left(\frac{dr^2}{1- \kappa r^2}+r^2 d\Omega^2\right)
\end{eqnarray}
forces the geometrical parameter $\kappa$ that appears in FLRW to coincide with the constant of integration $k$, used in the Newtonian Cosmology.

These results obtained in this paper
are of interest at least in two respects, one from the point of view
of its pedagogical value of teaching general relativity without in
fact using Einstein's equation and second, the fact that some results
attributed to general relativity can be obtained without using general
relativity indicates that these results are more general than the
particular dynamics specified by general relativity. Although,
some generalizations are possible. like the possibility of introducing a cosmological
constant in Newtonian Cosmology, as discussed in Bondi’s book \cite{4}
, we could in this way that by matching this cosmology to an exterior stationary space we obtain Schwarzschild- deSitter space, but Einstein's Equation does not expect that an approach of this type will be able to give all
results of general relativity, for example, certainly not in the case of gravitational waves or the Kerr solutions. For the case of inhomogeneous dust ball distribution, the CM lies on arbitrary origin and hence the linear and angular momentum remains zero. So also has been shown that the effective force between the constituent particles are zero for no perturbation \cite{5}.If we ponder minutely, the paper \cite{7}  goes very far just using Newtonian physics, but is missing a space-time interpretation of the solutions of Newton’s laws for gravitational interacting particles. We might try to find them properly even in presence of pressure in our next venture.

\medskip

 Finally an interesting question arises that whether our derivation holds only in the weak field approximation or not. Notice that indeed in parts of our arguments we have used
the weak field approximation, like when we derived the $0-0$ component
of the metric, then on this basis, we derived the $r-r$ component
by a process of matching to the internal collapsing ball of dust.
But if we take the point of view that we trust the metric of the collapsing
dust beyond the weak field approximation, the situation will be different,
in this case our derivation could have validity beyond the weak field
approximation, this is a question to be studied. One should also point
out that in general relativity, as far as the post-Newtonian approximation is concerned,
the corrections to the the $0-0$ component of the metric appear at
the same other as the first corrections to the $r-r$ components,
so it is in a sense puzzling that the $r-r$ component has been more
elusive to find by a simple derivation, as compared to the case of
the $0-0$ component.

\section{Appendix}

\subsection{\bf{Sommerfeld's approaches to this problem}}

There  had been some attempts of deriving the Schwarzschild metric invoking elementary considerations, like that explained in the book by Sommerfeld \cite{1} and attributed to Lenz where the fact that $g_{00} = −g^{-1}_{rr}$ is interpreted as a similar effect to the time dilatation and the space contraction. Such attempts have been criticized and the general consensus is that there cannot have a derivation of the
Schwarzschild metric along these lines \cite{2}.There exist although a reasonable elementary justification for the $g_{00}$ component of the Schwarzschild metric based on the Newtonian limit. An
interesting heuristic approach ( not a derivation ) however to obtain Einstein's Equation results has been
discussed by A. Rabinowitz\cite{3}. Another interesting heuristic approach to motivate (but not to
derive) the Schwarzschild metric include that of Visser \cite{4}
, which uses the intuition developed from analog gravity.

\medskip
It would be rather interesting to see how far one can go in the derivation of general relativity results without using in fact the Einstein's Equation at all. Sommerfeld \cite{1} tried to derive some aspects starting from the Schwarschild metric 

\begin{equation}
ds^{2}=-\gamma^{-2}dt^{2}+\gamma^{2}dr^{2}+r^{2}(d\theta^{2}+sin^{2}\theta d\phi^{2})
\end{equation} 
with $\gamma=(1-\frac{2M}{r})^{-1/2}$. He then conceived the idea that time-dilation and length contraction are actually related with the metric as follows:

\begin{equation}
dt_{0}=\gamma^{-1}dt
\end{equation}

\begin{equation}
dr_{0}=\gamma dr
\end{equation}

The time metric and space metric has been shown to be related with opposite powers of $\gamma$. But this construction has been shown to have many flaws \cite{8} \cite{9}. Instead of pushing ideas along the lines of Sommerfeld, we have investigated this problem in another direction which reader can go through in the main article.
 
\subsection{\bf{Friedmann\textendash Lemaître\textendash Robertson\textendash Walker
(FLRW) metric}}

\medskip

In Cosmology, the generalization to Einstein's static model of Universe
is FLRW model. This is isotropic and homogeneous model of universe
with the introduction of concept of curvature.

The metric equation reads as 

\begin{equation}
ds^{2}=-dt^{2}+a^{2}(t)[dr^{2}+f_{k}^{2}d\Omega^{2}]
\end{equation}

Where, 
\begin{equation}
d\Omega^{2}=d\theta^{2}+sin^{2}\theta d\phi^{2}
\end{equation}

$f_{k}$ is called the curvature function and is defined as below
:
\begin{equation}
f_{k}(r)= 
\begin{array} {c}k^{-1}sin(r\sqrt{k})\\
r\\
k^{-1}sinh(r\sqrt{k})
\end{array}
\begin{array}{c}
k>0\\
k=0\\
k<0
\end{array}
\end{equation}

here, k is called the curvature constant.The radius of curvature being
related to k as $r_{curvature}=\sqrt{k}.$ It is interesting to point out the physical meaning of the constant k. In standard values of k, as we know

\medskip
-For k = +1, positively curved and ” Closed " Universe.\\
-For k = 0, the space geometry is Euclidean or ” flat ” Universe\\
-For k = 1 , negatively curved, Space is infinite or ” open ” universe.\\

\medskip
With this the metric takes its standard form :

\begin{equation}
ds^{2}=-dt^{2}+[\frac{dr^{2}}{1-kr^{2}}+r^{2}d\Omega^{2}]
\end{equation}

Sometimes, it is defined in an alternate form

\begin{equation}
ds^{2}=-dt^{2}+a^{2}(t)\gamma_{ab}dx^{a}dx^{b}\;\forall a,b,c=1(1)3
\end{equation}

where $\gamma_{ab}$ can be expressed in the following form :

\begin{equation}
\gamma_{ab}=\delta_{ab}+\frac{kx_{a}x_{b}}{1-kx_{c}x^{c}}
\end{equation}

\subsection{\bf{ Fluid equation in presence of Pressure ($P \neq 0$)
}}

In this subsection we are going to find out the Fluid equation in presence of Pressure in the dust.
We are hereby going to consider the universe to be a sphere, where
exterior have no interaction with the matter inside. We can see the
volume is actually 

\begin{equation}
V(t)=\frac{4}{3}\pi r^{3}(t)=\frac{4}{3}\pi R^{3}(t)r^{3}(t_{0})
\end{equation}

So the time derivative will be 

\begin{equation}
\dot{V}(t)=4\pi R^{2}(t)\dot{R}(t)r^{3}(t_{0})
\end{equation}

Now Energy content

\begin{equation}
U=m=\frac{4}{3}\pi R^{3}(t)r^{3}(t_{0})\rho
\end{equation}

time derivative of this Energy 

\begin{equation}
\dot{U}=\frac{4}{3}\pi R^{3}(t)r^{3}(t_{0})\dot{\rho}+4\pi R^{2}(t)r^{3}(t_{0})\rho\dot{R}
\end{equation}

Since External environment has no interaction on the matter, so its
an adiabatic and isentropic process (with $dQ=dS=0$). With this condition
Second law of thermodynamics reads as follows:

\begin{equation}
\dot{U}+P\dot{V}=0
\end{equation}

\begin{eqnarray}
\Rightarrow\frac{4}{3}\pi R^{3}(t)r^{3}(t_{0})\dot{\rho}+4\pi R^{2}(t)r^{3}(t_{0})\rho\dot{R}+P4\pi R^{2}(t)\dot{R}(t)r^{3}(t_{0}) & = & 0
\end{eqnarray}

\begin{equation}
\Rightarrow\dot{\rho}+3\frac{\dot{R}}{R}(P+\rho)=0
\end{equation}

Which is renowned Fluid equation in presence of pressure.
\\

\section{Acknowledgment}
E.G. wants to thank the Fresno California State University Physics Department for the opportunity to present a preliminary version of the results of this paper in a colloquium. We wish to thank Gilad Granit, Tomer Ygael and Christian Rohrhofer who participated in an earlier version of this work. We also want to thank FQXI for financial support.

\end{document}